\documentclass[12pt]{article}

\def\beq{\begin{equation}}
\def\eeq{\end{equation}}
\def\barr{\begin{eqnarray}}
\def\beqa{\begin{eqnarray}}
\def\earr{\end{eqnarray}}
\def\eeqa{\end{eqnarray}}
\def\winf{W_{1+\infty}\ }
\def\u1{\widehat{U(1)}}
\def\v{V\,}
\def\w{W\,}
\def\vb{{\overline V}\,}
\def\wb{{\overline W}\,}

\newcommand{\nl}{\nonumber \\}

\begin{document}

\begin{titlepage}

\begin{center}
\hfill  \quad  \\
\vskip 0.5 cm
{\Large \bf Quantum dynamics of the effective field theory of the Calogero-Sutherland model }

\vspace{0.5cm}

Federico~L.~ BOTTESI$^a$ ,\ \ Guillermo~R.~ZEMBA$^{b,c,}$\footnote{
Fellow of Consejo Nacional de Investigaciones Cient\'{\i}ficas y T\'ecnicas, Argentina.}\\
\medskip
{\em $^a$Facultad de Ingenier\'ia, Universidad de Buenos Aires,}\\
{\em  Av. Paseo Col\'on 850,(C1063ACL) Buenos Aires, Argentina}\\
\medskip
{\em $^b$Departamento de F\'{\i}sica Te\'orica, GIyA, Laboratorio Tandar,}\\
{\em  Comisi\'on Nacional de Energ\'{\i}a At\'omica,} \\
{\em Av.Libertador 8250,(C1429BNP) Buenos Aires, Argentina}\\
\medskip
{\em $^c$Facultad de Ingenier\'ia y Ciencias Agrarias,  Pontificia Universidad Cat\'olica Argentina,}\\
{\em  Av. Alicia Moreau de Justo 1500,(C1107AAZ) Buenos Aires, Argentina}\\

\medskip

\end{center}
\vspace{.3cm}
\begin{abstract}
\noindent
We consider the known effective field theory of the Calogero-Sutherland model in the 
thermodynamic limit of large number of particles, obtained from the standard 
procedure in conformal field theory:
the Hilbert space is constructed {\it a priori} in terms of irreducible representations of the symmetry algebra,
and not by diagonalization of the hamiltonian, which is given in  
terms of fields that carry representations of the $\winf$ algebra (representing the incompressibility of the Fermi 
sea). Nevertheless, the role of the effective hamiltonian of the theory is to establish a specific dynamics,
which deserves further consideration.
We show that the time evolution of the (chiral or antichiral) density field is given by the quantum Benjamin-Ono equation, 
in agreement with previous results obtained from the alternative description of the continuous 
limit of the model, based on quantum hydrodynamics.
In this study, all calculations are performed at the quantum operator level, without making 
any assumption on the  semiclassical limit of the fields and their equations of motion. 
This result may be considered as a reliable indication of the equivalence between the quantum field theoretic and
quantum hydrodynamical formulations of the effective theories of the model.
A one-dimensional quantum compressible fluid that includes both chiralities is the physical picture that emerges 
for the continuous limit of the Calogero-Sutherland model.
\end{abstract}
\vskip 0.5 cm
\end{titlepage}
\pagenumbering{arabic}
\section{Introduction} 

The Calogero-Sutherland (CS) model \cite{cal,sut}
has a long and rich history in theoretical physics. 
It exhibits several interesting properties in its first 
quantized formulation, with a finite number of non-relativistic 
particles, as well as in its field theoretic formulation,
which describes the thermodynamic limit of infinite number 
of particles. 
In this work, we shall consider the second alternative,
after recalling that many of the interesting features of the
first one are reviewed, {\it e.g.}, in \cite{cs-reviews}.
The second quantized formulation of this model has been
studied with at least two types of formulations: 
the quantum hydrodynamic approach of \cite{aw,six} and the extended conformal field theory (CFT) \cite{bpz} 
procedure of \cite{clz,cfslz,flsz} (for alternative formulations, see also \cite{khve,amos}).
Here we shall consider the second alternative with the scope of 
developing a better understanding of the physical picture that is implied by it, and to make contact with 
the most relevant results and conclusions of the first option, which involves a semiclassical 
and attractive picture of the dynamics. 
Quantum hydrodynamics and field theory are two ways of studying quantum
matter and there relationship deserves to be more deeply understood (see, {\it e.g.}, \cite{jackiw}).

The extended CFT of the thermodynamic limit of the CS model 
naturally incorporates the idea of bosonization of the
lowest energy fluctuations of the Fermi surface as the relevant
semiclassical degrees of freedom \cite{boson} (see also \cite{kaya}). 
The same theory describes general Luttinger systems as well \cite{voit}.
The effective field theory (EFT) is constructed following the general prescription 
described in \cite{polch}:
the low-lying deformations of the Fermi surface are parametrized 
in terms of the generators of the $\winf$ dynamical symmetry \cite{shen,kac1,wref}, 
which, therefore, describe the lowest energy (``gapless'') fluctuations of the many-body states of the system
in an algebraic formulation. Furthermore, the EFT allows for the inclusion of non perturbative 
effects due to the 
interaction among fermions by a change in the representation of the symmetry algebra,
a result implied by Luttinger's theorem \cite{hald1}. This is the 
main reason for writing the EFT in terms of the $\winf$ generators,
which may be considered as favorable choice given the structure provided by this
mathematical structure.
In this work, we further investigate the physical consequences 
of this algebraic formulation of the effective theory,
focusing on the time evolution of the quantum density field.

This paper is organized as follows: in section 2, we review the EFT of the CS model
and the $\winf$ algebra. In section 3 we discuss the spectrum of the effective hamiltonian.
In section 4 we study the equation of time evolution of the density field and discuss 
the relationship of our results to those obtained by the quantum hydrodynamical formulation
of the CS effective theory. Finally, we provide some conclusions in section 5.

\section{The EFT of the Calogero-Sutherland model} 

We start by briefly reviewing the CS model. Consider a system of N non-relativistic 
$(1+1)$-dimensional spinless interacting fermions on a circle of length L, with 
hamiltonian \cite{sut} (in units where $\hbar =1$ and $2m=1$ , with $m$ being 
the mass of the particles)
\beq
h_{CS}=\sum_{j=1}^N\ \left( \frac{1}{i} \frac{\partial}{\partial x_j}\
\right)^2\ +\ g\ \frac{\pi^2}{L^2}\ \sum_{i<j}\ \frac{1}{\sin^2
(\pi(x_i-x_j)/L) }\ ,
\label{ham}
\eeq
where $x_i$ ($i=1,\dots,N$) is the coordinate of the $i$-th particle, and $g$ is 
the dimensionless coupling constant. Ground state stability demands $g \geq -1/2$,
with both attractive ($-1/2 \leq g < 0$) and repulsive ($0 < g$) regimes. A usual 
reparametrization of the coupling constant is given by $g=2 \xi ( \xi -1)$, 
so that $\xi \geq 0$ and $0 \leq \xi < 1$ is the attractive regime and $1  < \xi $ 
the repulsive one. 

The $\winf$ EFT of (\ref{ham}) has been derived in 
\cite{clz}\cite{cfslz}\ by formulating the system dynamics in terms of fields describing the fluctuations of the 
$1D$ Fermi surface in the thermodynamic limit. 
This method amounts to initially defining suitable non-relativistic fermionic fields, 
taking then the limit $N \to \infty$ properly on them and the hamiltonian 
obtaining an EFT involving two sets of independent relativistic fermion fields 
that describe the low energy fluctuations 
around each of the two Fermi points of the $1D$ effective theory. 
The term relativistic here means a linear dispersion relation. The last 
step consists in writing down the EFT in terms of fields that obey the $\winf$ 
symmetry. This step is crucial in our approach, as it allows to diagonalize the 
hamiltonian and find the Hilbert space of the EFT by exploiting the algebraic 
properties.
Moreover, the $\winf$ algebra found in the original basis of fermion bilinears may be realized as well by
bosonic operators, that take into account interactions in the fermionic picture, 
as anticipated.
Here we outline the procedure: in the free case ($g=0$), the single-particle wave functions 
are given by plane waves:
\beq
\phi_k (x)\ = \frac{1}{\sqrt{L}}\ \exp\ \left( i \frac{2\pi}{L} k x
\right)\ ,
\label{wfcir}
\eeq
where $k$ is an integer for periodic boundary conditions.
The new variables are derived from a second quantized non-relativistic fermion field 
defined in terms of the above wave functions as:
\beq
\Psi (x,t)\ \equiv\ \sum_{k=-\infty}^{\infty}\ c_k\  \phi_k (x,t)\  ,
\qquad \{\ c_k , c^{\dag}_l\ \}\ =\ \delta_{k,l}\ .
\label{field}
\eeq
Here $\phi_k (x,t)=\phi_k (x)\exp\left(-i {\epsilon}_k t \right ) $,
${\epsilon}_k = (2\pi/L)^2 k^2$
and $c_k$, $c^{\dag}_l$ are Fock space operators.
The ground state of the system is
\beq
|\ \Omega\ , N\ \rangle\ =\ c^{\dag}_{-M} c^{\dag}_{-M+1}\dots
c^{\dag}_{M-1}c^{\dag}_{M} |\ 0\ \rangle\ ,
\label{vac}
\eeq
where $|\ 0\ \rangle$ is the Fock vacuum and $M \equiv (N-1)/2$.
The ground state particle number density $n(x,t)$ is given by
\beq
n(x,t)\ \equiv\
\langle\ \Omega\ , N\ |\Psi^{\dag} (x,t) \Psi (x,t)
|\ \Omega\ , N\ \rangle\ =\ \sum_{k=-M}^{M}
|\phi_k(x,t) |^2\ =\ \frac{N}{L}\ .
\label{gsd}
\eeq
Therefore the spatial density $n(x,t)=n_0=N/L$ is uniform and stationary.
Next we consider the thermodynamic limit of large $N$ and $L$,
with $n_0$ finite. The Fermi sea a segment in momentum space between the two Fermi points,
located at $\pm p_F$, with the Fermi momentum $p_F=\pi(N-1)/L$.  
We refer to these two Fermi points by their locations as $R$ and  $L$, for right 
and left, respectively (we hope that no confusion arises with the length variable).
The ground state
(\ref{vac}) becomes a relativistic Dirac sea for each one of them, in this limit.
The new variables 
are defined as `shifted' Fock operators around each Fermi point
\cite{boson,hald1,cdtz}: $a_r\ \equiv\ c_{M+r}\ $ for the $R$ Fermi point and
${\overline a}_r\ \equiv\ c_{-M-r}\ $ for the $L$ one , where $ |r| \ll \sqrt{N}$ 
describes small fluctuations, {\it i.e.}, the lowest energy excitations of the system.
The fields become Weyl fermions \cite{bpz}, describing the relevant degrees of freedom
of the system in the vicinity of the Fermi points \cite{polch}. The
corresponding hamiltonian may be written 
in terms of new bilinear fermionic fields that satisfy the $\winf$ algebra. 
Once this form is found for free fermions, a one parameter deformation keeping 
the algebraic structure extends this expression to a bosonized 
form of the theory.

Making use of the bosonization expressions derived in \cite{clz,cfslz}, the 
effective hamiltonian becomes the following operator: 
\barr
{\cal H}_{CS} &=&
\left\{ \ \left(2\pi n_0  \sqrt{\xi}\right)^2\ 
\left[\frac{\sqrt{\xi}}{4}\,\w_0^0
+\frac{1}{N}\,\w_0^1+
\frac{1}{N^2}\left(\frac{1}{\sqrt{\xi}}\,\w_0^2
-\frac{\sqrt{\xi}}{12}\,\w_0^0 \right.\right.
\right.\nl
&&-\ \left.\left.\left.
\frac{g}{2\xi^2}\,\sum_{\ell=1}^\infty
\,\ell~\w_{-\ell}^0\,\w_\ell^0\right)
\right]+\left(\,W~\leftrightarrow~{\overline W}\,\right)
\right\}~~~,
\label{hcsf}
\earr
where $\xi=\left(1+\sqrt{1+2g}\right)/2$ is the parameter defined after (\ref{ham}).
Note that $\xi = 1$ corresponds to the free fermion case.
Nevertheless, the relationship between $\xi $ and $g$ is not relevant when discussing 
the properties of the EFT, which leaves behind all the small scale details 
of the underlying dynamics. Therefore, 
$\xi$ is assumed to be a real positive {\it free independent} parameter for the rest of the discussion.
The operators $~\w_{\ell}^m$ in (\ref{hcsf}) are the lowest generators of the $\winf$ infinite dimensional algebra 
\cite{shen,kac1}. Here $m=i+1$, where $i$ is the conformal spin \cite{bpz}.
 The terms in the $~\w_{\ell}^m$ ($\wb^i_\ell$) operators describe the 
dynamics at the right $(R)$ (left $(L)$) Fermi point, respectively. 
We remark that the {\it complete factorization} of (\ref{hcsf}) into chiral
and antichiral sectors is the consequence of rotating the fields 
$~\v_{\ell}^m$ and $\vb^i_\ell$ by a bosonic Bogoliubov transformation.
These sectors are mixed 
by backward scattering terms in the first fermionic form of the hamiltonian obtained 
\cite{cfslz,flsz}.

The general form of the $\winf$ algebra is:
\beq
\left[\ \w^i_\ell, \w^j_m\ \right] = (j\ell-im) \w^{i+j-1}_{\ell+m}
+q(i,j,\ell,m)\w^{i+j-3}_{\ell+m}
+\cdots +\delta^{ij}\delta_{\ell+m,0}\ c\ d(i,\ell) \ ,
\label{walg}
\eeq
where the structure constants $q(i,j,\ell,m)$ and $d(i,\ell)$ 
are polynomial in their arguments, $c$ is the central charge, 
and the dots denote a {\it finite} number of terms involving the operators 
$\w^{i+j-2k}_{\ell+m}\ $.
The ground state $|\ \Omega\ \rangle$ is a highest weight state 
with respect to the $\winf $ operators, namely
$\w^i_\ell |\ \Omega\ \rangle = 0$, $\ell > 0, i \geq 0$,
that is to say that it is incompressible in momentum space. 
We remark here that the basis $\w^i_\ell$ of $\winf$ operators in 
the hamiltonian (\ref{hcsf}) is bosonic and not the original fermionic one
denoted as $\v^i_\ell$, inherited from the 
CS model, as we shall explain right away. 
The major advantage of choosing the basis of the $\winf \times 
{\overline \winf}$ operators is that, once the algebraic content 
of the theory has been established in the free fermionic picture,
the bosonic realization of the algebra can be used,
and the free value of the compactification radius of the boson
can be chosen so as to diagonalize the hamiltonian.
This method is consequently termed as {\it algebraic bosonization} \cite{flsz}.
For the case of the CS EFT $c=1$, $\winf$ is interpreted as the enveloping
algebra of the fermion number $U(1)$ symmetry, and all the relevant commutation relations are:
\barr
\left[\ \w^0_\ell,\w^0_m\ \right] & = &  c\ \xi \ell\ \delta_{\ell+m,0} ~~~,\nl
\left[\ \w^1_\ell, \w^0_m\ \right] & = & -m\ \w^0_{\ell+m} ~~~,\nl
\left[\ \w^1_\ell, \w^1_m\ \right] & = & (\ell-m)\w^1_{\ell+m} + 
\frac{c}{12}\ell(\ell^2-1) \delta_{\ell+m,0}~~~,\nl
\left[\ \w^2_\ell, \w^0_m\ \right] &=& -2m\ \w^1_{\ell+m}~~~,
\label{walg1}\\
\left[\ \w^2_\ell, \w^1_m\ \right] &=& (\ell-2m)\ \w^2_{\ell+m} -
   \frac{1}{6}\left(m^3-m\right) \w^0_{\ell+m}~~~,\nl
\left[\ \w^2_n, \w^2_m\ \right] &=& (2n-2m)\ \w^3_{n+m}
     +{n-m\over 15}\left( 2n^2 +2m^2 -nm-8 \right) \w^1_{n+m}\nonumber\\
     &&\quad +\ c\ {n(n^2-1)(n^2-4)\over 180}\ \delta_{n+m,0}~~~.\nonumber
\earr
The first and third equations in (\ref{walg1}) show
that the generators $\w^0_\ell$ satisfy the 
abelian Kac-Moody algebra $\u1$, and the generators $\w^1_\ell$ 
satisfy the Virasoro algebra, respectively.
The operators $\wb^i_\ell$ fulfil the same algebra (\ref{walg1}) 
with central charge ${\overline c}=1$ 
and commute with the all the operators $\w^i_\ell$. The complete EFT 
of the CS model is a $(c,{\overline c})=(1,1)$ CFT, but since both chiral ($R$) and
antichiral ($L$) sectors are isomorphic, we will often consider one of them at a time, for
the sake of simplicity.

The $c=1$ $\winf$ algebra can be realized by either {\it fermionic} or {\it bosonic} 
operators. In our methodology for constructing EFTs, the first realization is 
useful for identifying the correct hamiltonian 
terms if one starts from fermionic systems, like the CS model,
as it was mentioned above. Its explicit form in terms of fermion bilinears is give by:
\barr
\v^0_n &=& \sum_{r=-\infty}^{\infty}: a^\dagger_{r-n}\, a_r :
{}~~~,\nl
\v^1_n &=& \sum_{r=-\infty}^{\infty} \left(\,r-{n+1\over 2}\,\right)
:  a^\dagger_{r-n}\, a_r :~~~,\label{fockw}\\
\v^2_n &=& \sum_{r=-\infty}^{\infty} \left(\,r^2 -(n+1)\ r +
{{(n+1)(n+2)}\over 6}\,
\right) :  a^\dagger_{r-n}\, a_r :~~~,\nl
\nonumber
\earr
with $\{\ a_k , a^{\dag}_l\ \}\ =\ \delta_{k,l}\ $ and all other 
anticommutators vanishing. These are explicit expressions for the $R$
Fermi point with analogous ones for the $L$ one.
The $\winf$ algebra satisfied by these operators is isomorphic to
(\ref{walg1}) with $\xi = 1$.

The second realization may be derived through a generalized Sugawara 
construction \cite{kac1} in terms of a chiral bosonic field.
In fact, if one introduces the right and left moving modes,
$\alpha_\ell$ and ${\overline \alpha}_\ell$, of a free compactified 
boson ($[\alpha_n,\alpha_m ]=\xi n \delta_{n+m,0}$ and similarly for 
the ${\overline \alpha}_\ell$ 
operators), one can check that the commutation relations (\ref{walg1}) 
are satisfied by defining $\w^i_\ell$ (we only write the expressions
for $i=0,1,2$) as
\barr
\w^0_\ell &=& \alpha_\ell ~~~,
\nonumber\\
\w^1_\ell &=& {\frac{1}{2}} \sum_{r= -\infty}^{\infty}
:\, \alpha_{r}\,\alpha_{\ell-r}\,
:~~~,\label{mod2}\\
\w^2_\ell &=& {\frac{1}{3}} \sum_{r, s = -\infty}^{\infty}
:\, \alpha_{r}\,\alpha_s\, \alpha_{\ell-r-s}\,:~~~,
\nonumber
\earr
and analogously for the operators $\wb^i_\ell$ in terms of ${\overline \alpha}_\ell$.
The naive generalization of (\ref{mod2}) to higher 
values of conformal spin is incorrect (for example, see \cite{flsz}).
Finally, the relationship between these two realizations is 
given by the correspondence $ \w^0_\ell \leftrightarrow \sqrt{\xi} \v^0_\ell$ and 
$ \w^i_\ell \leftrightarrow \v^i_\ell$ ($i=1,2,\dots $). It follows that the
constant density is renormalized by a 
factor $1/ \xi $ with respect to the free fermion case.

A remark about the hamiltonian (\ref{hcsf}) is that it is
given as a power series expansion in the small parameter $1/N$,
but a {\it finite} one: the expansion stops at $1/N^2$. 
This result agrees with \cite{aw} 
and reflects the fact that the dispersion relation at the Fermi points is not linear,
but still polynomial (quadratic in this case): on the contrary, 
the same method applied to the Heisenberg model yields an infinite power 
series \cite{flsz}. For the CS EFT, it is a consequence of the dimensionality of the interaction 
(\ref{ham}) that decays as $1/L^2$.
Note that the terms of $O(1)$ in (\ref{hcsf}) correspond to global operators (zero modes), the 
terms of $O(1/N)$ are the finite-size universal corrections given by CFT and the 
terms of $O(1/N^2)$ are beyond the usual scope of CFT.

\section{Spectrum of the effective hamiltonian} 

A key remark regarding the derivation of (\ref{hcsf}) in \cite{clz,cfslz},
is that the hamiltonian may be written as a sum of decoupled chiral sectors 
only after a Bogoliubov transformation has been performed on the fermionic 
$\winf$ operators. Indeed, the effective hamiltonian is naturally written in terms
of the fermionic basis of the $\winf$ operators (\ref{fockw}), and has backward
scattering terms (that mix both chiralities). The Bogoliubov transformation
decouples both Fermi points and defines a new $\winf$ basis that, moreover, is rewritten in 
the bosonic form (\ref{mod2}), is given by:
\barr
\w^0_{\ell}&=&\v^0_{\ell}\ \cosh \beta + \vb^0_{-\ell}\
\sinh \beta ~~~, \nl
\wb^0_{\ell}&=&\v^0_{-\ell}\ \sinh \beta +
\vb^0_{\ell}\ \cosh \beta
\label{bogo}
\earr
for all $\ell$, with
\beq
 \tanh 2\beta =\frac{\xi -1}{\xi + 1} ~~~.
\label{angle}
\eeq
Equivalently, $\exp (2\beta) = \xi $ (see \cite{flsz,boze3} for further details).

We now discuss the spectrum of
(\ref{hcsf}). In the fermionic description 
$\v^i_\ell$ it is easy to see that the highest 
weight states of the $\winf \times {\overline \winf}$ algebra
are attained by the addition of $\Delta N$ particles to the 
ground state $|\Omega \rangle$, and by moving $\Delta D$ 
particles from the left to the right Fermi point; they are 
denoted by $|\Delta N,\Delta D\rangle_0$.
The descendant states, 
$$
|\Delta N , \Delta D ; \{k_i\},\{{\overline k}_j\} 
\rangle_0 \ = 
\v^0_{-k_1} \dots \v^0_{-k_r} \vb^0_{-{\overline k}_1}
\dots \vb^0_{-{\overline k}_s}
|\Delta N , \Delta D \rangle_0~~~~,
$$
with $k_1 \ge k_2 \ge \dots \ge k_r > 0$, and
${\overline k}_1 \ge {\overline k}_2 \ge \dots 
\ge {\overline k}_s > 0$,
coincide with the particle-hole excitations arising from
$|\Delta N , \Delta D \rangle_0$.
Using the expressions of $\v_0^0$ and $\vb_0^0$ given in
(\ref{fockw}), one finds that the charges associated to 
these states are
\barr
\v_0^0 ~|\Delta N , \Delta D ; \{k_i\},\{{\overline k}_j\} 
\rangle_0&=& \left(\frac{\Delta N}{2} \ + \Delta D  \right)
|\Delta N , \Delta D ; \{k_i\},\{{\overline k}_j\} \rangle_0~~~,
\nl
\vb_0^0 ~ |\Delta N , \Delta D ; \{k_i\},\{{\overline k}_j\} 
\rangle_0 \ &=& \left(\frac{\Delta N}{2} \ - \Delta D  \right)
|\Delta N , \Delta D ; \{k_i\},\{{\overline k}_j\} \rangle_0~~~.
\label{deltandeltad}
\earr
In terms of the bosonized operators basis $\w^i_\ell$  the highest weight vectors,
$|\Delta N ; \Delta D \rangle_W$, are still
characterized by the numbers $\Delta N$ and $\Delta D$ with the
same meaning as before, but their charges are different.
More precisely
\barr
\w_0^0 ~|\Delta N ; \Delta D \rangle_W  &=&
\left(\sqrt{\xi}\,\frac{\Delta N}{2}+
\frac{\Delta D}{\sqrt{\xi}}\right)
|\Delta N ; \Delta D \rangle_W \nl
\wb_0^0 ~|\Delta N ; \Delta D \rangle_W  &=&
\left(\sqrt{\xi}\,\frac{\Delta N}{2}-
\frac{\Delta D}{\sqrt{\xi}}\right)
|\Delta N ; \Delta D \rangle_W~~~.
\label{vdo}
\earr
The highest weight states $|\Delta N , \Delta D \rangle_W$
together with their descendants, denoted by
$|\Delta N , \Delta D ; \{k_i\},\{{\overline k}_j\} \rangle_W$,
form a {\it new} bosonic basis for the
theory that has no simple expression in
terms of the original free fermionic degrees of freedom.

The exact energies of these excitations in this basis are given by:
\barr
{\cal E}&=&
\left\{ \ \left(2\pi n_0  \sqrt{\xi}\right)^2\ 
\left[\frac{\sqrt{\xi}}{4}\,Q+\frac{1}{N}
\left(\frac{1}{2}\,Q^2+k\right)
+\frac{1}{N^2}\left(\frac{1}{3\sqrt{\xi}}\,Q^3
-\frac{\sqrt{\xi}}{12}\,Q\right.\right.\right.
\label{eba} \\
&&+\ \left.\left.
\frac{2k}{\sqrt{\xi}}\,Q + \frac{\sum_j k_j^2}{\xi}
-\sum_j \left(2j-1\right) k_j\right)\right]
+\left(Q\ \leftrightarrow \ {\overline Q}~,
{}~\{k_j\} \ \leftrightarrow \ \{{\overline k}_j\} \right)\Bigg\}~~~,
\nonumber
\label{exacte}
\earr
where
$$ k \ =\ \sum_j k_j ~~~~,~~~~
{\overline k} \ =\ \sum_j {\overline k}_j
$$
and the eigenvalues of $\w_0^0$ and $\wb_0^0$, respectively, are
\beq
Q=\sqrt{\xi}\,\frac{\Delta N}{2}+
\frac{\Delta D}{\sqrt{\xi}}~~~~,~~~~
{\overline Q}=\sqrt{\xi}\,\frac{\Delta N}{2}-
\frac{\Delta D}{\sqrt{\xi}}~~~.
\label{Q}
\eeq
Moreover, the integers $k_j$ are ordered
according to $k_1\geq k_2\geq \dots \geq 0$,
and are different from zero only if $j << \sqrt{N}$, {\it i.e.}, within the range of validity of the 
EFT, and analogously for ${\overline k}_j$.
Notice that $Q$ and ${\overline Q}$ have
the structure of the zero mode charges of a non-chiral
bosonic field (which is the sum of chiral and antichiral bosons) 
compactified on a circle of radius
\beq
r = \frac{1}{\sqrt{\xi}}~~~.
\label{rtil}
\eeq
Indeed, this is the exact value of the compactification
radius of the bosonic field describing the density fluctuations
of the fermions in the CS model \cite{kaya}.
The partition function of this $c=1$ CFT is known to be invariant 
under the duality symmetry $r \leftrightarrow 1/(2r)$, which in our language 
is equivalent to the mapping $\sqrt{\xi} \leftrightarrow 2/\sqrt{\xi}$.
The action of this mapping on the charges (\ref{Q}) is to interchange 
$\Delta N$ with $\Delta D$, such that $Q$ remains unchanged and ${\overline Q}$
maps onto minus itself. Some known identifications are: $\xi = 2$  
is the self-dual point,  $\xi = 1$  the free fermion point and $\xi = 1/2$ 
the Kosterlitz-Thouless point \cite{bpz}.
Notice that the zero modes are the only links between the $L$ and $R$
Fermi points within the framework of the EFT.
From (\ref{eba}) we can also see that the exact value of the
Fermi velocity is
\beq
v = 2\pi n_0\ \xi~~~,
\label{vti}
\eeq
as a consequence of Luttinger's theorem , that demands that the product 
$v r^2 $ remains constant for any value of $\xi$ \cite{hald1}.
In more physical terms, the spectrum (\ref{exacte}) corresponds to both charged and neutral
low-lying excitations. The charged (with respect to the $U(1)$ symmetry of fermion number) ones are
labeled by $Q$, which is interpreted as a {\it soliton number}, represented in CFT 
by local vertex operators \cite{bpz}. The 
neutral excitations are given by the integers $k_j$ 
and correspond to {\it particle-hole}-like excitations. This analysis stems 
from the fermionic picture and generalizes to the bosonic one \cite{clz}.
The structure of the spectrum is familiar in CFT: the charged excitations are 
highest weight states 
and the neutral fluctuations correspond to the Verma modules on top of each one 
of them \cite{bpz}.
That is to say that the EFT describes a uniform density ground state that may have 
solitons (located lumps or valleys) and fluctuations around them as the complete set of 
low-lying excitations.
Unusually for a CFT, we have higher order corrections in $Q$ beyond $Q^2$ in the hamiltonian,
because the universal finite-size corrections are of order $1/N$. 

\section{Dynamics of the density field operator} 

In $(1+1)$ systems with quantum dynamics, such as the CS model, fields and operators are naturally defined 
on the boundary circle $(0\le\theta <2\pi)$ where $x = R \theta$, {\it i.e.},
on a compact space \cite{wref}. In the mathematical literature, however, they are conventionally considered in 
an unbounded space.
There is a conformal mapping between these two spaces, which are the {\it cylinder}
$(u=\tau  -iR \theta)$ and the {\it conformal plane} $(z)$,  after inclusion of the Euclidean time $\tau $:
\beq
z\ =\ \exp\left(\frac{u}{R}\right)\ =\
\exp\left(\frac{\tau}{R} -i\theta \right) \ .
\label{confmap}
\eeq
where $x$ is periodic with period $L$: $x = R \theta$ with $R=L/(2 \pi )$. For example,
the operators (\ref{walg1}) define the $\winf$ currents $W^i(z)$
on the conformal plane as follows,
\beq
W^i(z)\ \equiv\ \sum_n\ W^i_n\ z^{-n-i-1}\ .
\label{fourv}
\eeq
When studying the dynamics of a physical system like the CS model, the hamiltonian of edge excitations should be
expressed in terms of the $\winf$ generators $(W_R)^i_n$ on the cylinder. 
\barr
W^0 (u)\ &=& \frac{dz}{du}\ W^0(z)\ , \cr
W^1 (u)\ &=& \left(\frac{dz}{du} \right)^2\ W^1(z)\ + \frac{c}{12}\
S(z,u)\ , \cr
W^2 (u)\ &=& \left(\frac{dz}{du} \right)^3\ W^2(z)\ + \frac{1}{6}\
\frac{dz}{du}\ S(z,u)\ W^0(z)\ . 
\label{wcyl}
\earr
The $\winf$ currents on the cylinder are thus found by using
the mapping ,
\barr
W^0_R (u)\ &=& \frac{z}{R} W^0(z)\ ,\cr
W^1_R (u)\ &=& \frac{z}{R^2}\left( z^2 W^1(z) - \frac{1}{24}\right)\ ,\cr
W^2_R (u)\ &=& \frac{z}{R^3}\left( z^3 W^2(z) -
\frac{z}{12}V^0(z)\right)\ .
\label{wcurc}
\earr
Using the definition
\beq
\left(W_R\right)^j_0\equiv\ \int _0^{2\pi iR} \frac{du}{(-2\pi i)}
\ W^j_R (u)\ ,
\label{dhw}
\eeq
we find the relation between the zero modes in the two geometries
\beq
\left(W_R\right)^0_0 = W^0_0\ ,\qquad
\left(W_R\right)^1_0 = \frac{1}{R} \left( W^1_0-\frac{c}{24} \right) \ ,
\qquad
\left(W_R\right)^2_0 = \frac{1}{R^2} \left( W^2_0
-\frac{1}{12} W^0_0 \right) \ .
\label{wgc}
\eeq
We now consider the transformation of the effective hamiltonian (\ref{hcsf})
from the plane to the cylinder geometry.
There is a natural unit of energy, with $\hbar = c =1$ given by:
\beq
E_{CS}\ =\ \frac{2{\pi}^2 n_0^2  \xi}{m}\ =\ \frac{ \pi n_0 \xi N}{mR}\ ,
\eeq
where we have momentarily restored the mass $m$ of the CS particles
for the purposes of comparison-
The dimensionless Fermi velocity is 
\beq
v\ =\ \frac{\pi n_0\ \xi}{m}\  =\ v_0\ \xi \ .
\label{vti2}
\eeq
The effective hamiltonian on the $(\tau,\theta)$ cylinder is, therefore, the following operator: 
\barr
H_{CS} &=&
\left\{ \ E_{CS}\ 
\left[\frac{\sqrt{\xi}}{4}\,\w_0^0
+\frac{1}{N}\,\w_0^1 +
\frac{1}{N^2}\left(\frac{1}{\sqrt{\xi}}\,\w_0^2
-\frac{\sqrt{\xi}}{12}\,\w_0^0 \right.\right.
\right.\nl
&&-\ \left.\left.\left.
\frac{( \xi -1)}{\xi}\,\sum_{\ell=1}^\infty
\,\ell~\w_{-\ell}^0\,\w_\ell^0\right)
\right]+\left(\,W~\leftrightarrow~{\overline W}\,\right)
\right\}~~~.
\label{hcsf2}
\earr
This is indeed defined on the cylinder because the hamiltonian does not involve terms with 
dimensional factors of powers of $1/R$. However, the Laurent modes are those 
defined by the commutator relations of the $\winf$ algebra (\ref{walg}) on the plane,
as usual.

We are now in a suitable position to investigate the dynamics induced by the effective hamiltonian (\ref{hcsf2}).
We choose to study the time evolution of the chiral component the {\it density field}, as it the most obviously related to the 
semiclassical hydrodynamical approach to the CS model. Note, however, that the hamiltonian
is not needed in order to determine the Hilbert space of the EFT, which is known since the beginning by the `kinematical
construction of CFTs. The antichiral sector of the theory could be considered as well along similar lines.
The density field equation of motion is related to that of the operator $W^0_R$:
\beq
\frac{\partial  W^0_R (u)}{\partial t}\ =\ -i\left[ W^0_R (u), H_{CS} \right]
=\ -i\frac{z}{R}\left[ W^0 (z), H_{CS} \right]
\label{hameqmo}
\eeq
We use the Sugawara construction with the normal ordering defined on the $z$ plane:
\beq
\w^1 ( z ) = \frac{1}{2}:\left( \w^0 ( z ) \right)^2 :\ ,
\eeq
Furthermore, we define the density field as
\beq
n(x,t)\ =\ \frac{1}{\pi \xi^{3/2}}\ \left( \w_R^0 (x,t) + \pi n_0 \sqrt{\xi}\ \right)\ ,
\label{dyndens}
\eeq
where $\w_R^0 (x,t) $ is the fluctuation field over the uniform value
$\langle \Omega |n(x,t) |\Omega \rangle = n_0 /{\xi}$, with
$\langle \Omega |\w_R^0 (x,t) |\Omega \rangle = 0 $.
We find:
\barr
\frac{\partial  n}{\partial t}\ &=&
\left( \frac{\pi  \xi^{2} }{2m}\ \right)\ 
\frac{\partial}{\partial x} \left(  n^2 
\right)\ +\
\nl
&&-\ 
\frac{iz( \xi -1)n_0}{m\sqrt{\xi}NR^2 }\,\sum_{\ell=1}^\infty
\,\left[\ell^2~z^{-\ell -1}\w_{\ell}^0\ -
\ell^2~z^{\ell -1}\w_{-\ell}^0\
\right]\ ~~~,
\label{hameqm10}
\earr
Note that the free fermion part coincides exactly with the {\it quantum Hopf equation} \cite{aw}.

Next we focus on the interaction term in (\ref{hameqm10}), which may be rewritten as:
\beq
\sum_{\ell=1}^\infty
\,\left[\ell^2~z^{-\ell -1}\w_{\ell}^0\ -
\ell^2~z^{\ell -1}\w_{-\ell}^0\ \right]\ =\ -\frac{\partial}{\partial z}
\left[
z \frac{\partial}{\partial z}\left( z W_{+}^0 (z)\right)\ 
-\ z \frac{\partial}{\partial z}\left( z W_{-}^0 (z)\right)\ 
\right]\ ,
\label{intew}
\eeq
where 
\barr
W_{+}^0 (z)\ &=& \sum_{n=1}^\infty\ W_{n}^0\ z^{-n-1}\ \nl
W_{-}^0 (z)\ &=& \sum_{n=-\infty}^{-1}\ W_{n}^0\ z^{-n-1}\ .
\label{halfw}
\earr
are the positive and negative Laurent mode fields. 
The equation of motion becomes: 
\barr
\frac{\partial  n}{\partial t}\ &=&
\left( \frac{\pi  \xi^{2} }{2m}\ \right)\ 
\frac{\partial}{\partial x} \left(  n^2 
\right)\ +\
\nl
&&-\ 
\frac{i( \xi -1)}{2\pi  m \sqrt{\xi} }\,
\frac{\partial^2}{\partial x^2}
\left(  \left( W_{R}^0 \right)_{+} \ - \  \left( W_{R}^0 \right)_{-} \right)\ .
\label{hameqm14}
\earr
We would like to understand better the non familiar field that appears in the interaction term. 
It corresponds to an interaction that in terms of the density field is local in space but non-local 
in the time domain. Indeed, according to \cite{aw}:
\beq
\left( W_{R}^0 \right)_H (x,t)\ =\ \left(  W_{+}^0 (u) \ - \  W_{-}^0 (u)\  \right) 
\label{wdecomp}
\eeq
where $\left( W_{R}^0 \right)_H (x,t) $ is the Hilbert transform of the field
$W_{R}^0 (x,t)$:
\beq
\left( W_{R}^0 \right)_H (x,t)\ =\ \frac{1}{\pi}
PV\ \int_{-\infty}^{\infty}\ \frac{W_{R}^0 (x,t') }{(t-t')}\ dt'\ ,
\label{defhilberttrans}
\eeq
where $PV$ denotes the Principal Value (there is a potential singularity
at $t' = t$). 
We now show that (\ref{wdecomp}) may be derived within the framework of CFT as well. We have 
\beq
W^0_R (u)\ = \frac{z}{R} W^0(z)\ =\ \frac{1}{R}\ \sum_{n=-\infty}^{\infty}\ 
z^{-n}\ W_{n}^{0}\ ,
\eeq
where $z=\exp\left[(\tau -i x)/R \right]$, $t = i\tau $ and 
$(\tau -i x)\ =\ R \ln z $. 
The integral in the $z$ plane is along the radial direction ('time evolution' on the plane).
\beq
\left( W_{R}^0 \right)_H (x,\tau)\ =\ \frac{1}{\pi}\ \sum_{n=-\infty}^{\infty}\ 
\ W_{n}^{0}\ 
PV\ \int_{0}^{\infty}\ 
\frac{(z')^{-n-1}}{(\ln{z}-\ln{z'})}\  dz'\ ,
\label{defhilberttrans5}
\eeq
where we assume $\theta'= \theta$, that is, spatial locality, and setting 
$|z|=r$ and $|z'|=r'$ , such that the potential singularity is now at $z'= z$, we have:
\beq
\left( W_{R}^0 \right)_H (x,\tau)\ =\ -\frac{1}{\pi}\ \sum_{n=-\infty}^{\infty}\ 
\ W_{n}^{0}\ 
PV\ \int_{0}^{\infty}\ \frac{(r')^{-n-1}}{\ln{r'/r}}\  dr'\ ,
\label{defhilberttrans7}
\eeq
setting $s=r'/r $, the potential singularity is now at $s= 1$.
Therefore, we arrive to the expansion
\beq
\left( W_{R}^0 \right)_H (x,\tau)\ =\ -\frac{1}{\pi}\ \sum_{n=-\infty}^{\infty}\ 
C_n\  \exp(-n\tau /R )\ W_{n}^{0}\ ,
\label{defhilberttrans9}
\eeq
with the coefficients 
\beq
C_n\ = PV\ \int_{0}^{\infty}\ \frac{s^{-n-1}}{\ln s}\  ds\ .
\eeq
These coefficients may be evaluated, taking care of subtracting the UV divergences properly, so that:
\begin{equation}
C_n = \left\lbrace
\begin{array}{ll}
-i\pi \qquad &{(n > 0)} \nl
+i\pi  \qquad   &{(n < 0)}
\end{array}
\right.
\label{coeffsn3}
\end{equation}
Therefore, the coefficients perform the projection on the positive and negative modes of the Laurent expansion 
of the field $W^{0}$.
Analogously, for the $0$-mode we find $C_0 = -i \pi $, so that
we can add it to the definition of $C_n$ with $n > 0 $. We have, therefore, shown that: 
\beq
\left( W_{R}^0 \right)_H \ =\ i \left[  \left( W_{R}^0 \right)_{+}\ -\ 
\left( W_{R}^0 \right)_{-} \right]\ .
\label{defhilberttrans12}
\eeq
Replacing this result in (\ref{hameqm14}) we finally obtain:
\beq
\frac{\partial  n}{\partial t}\ =\ 
\left( \frac{\pi  \xi^{2} }{2m}\ \right)\ 
\frac{\partial}{\partial x} \left(  n^2 
\right)\ +\ \frac{( \xi -1)\xi }{2  m}\,
\frac{\partial^2 n_H}{\partial x^2}\ ,
\label{hameqm16}
\eeq
where $n_H$ is the Hilbert transform of the density, defined by (\ref{dyndens}) with $W_{R}^0$
replaced by $\left( W_{R}^0 \right)_H$.
This is the quantum Benjamin-Ono (BO) equation \cite{aw} , written with two coupling constants. It is identical 
to the classical BO equation but it is satisfied by a quantum field rather than by a real function.
The classical Benjamin-Ono equation is a nonlinear partial integro-differential equation that describes 
one-dimensional internal waves in deep water \cite{benja,ono} .
Each term has dimension $2$ (in natural units),
therefore the couplings $\alpha = \pi  \xi^{2} /(2m) $ and 
$\beta =  \xi ( \xi -1) /(2  m) $ have both dimension $1$. Their 
quotient is, therefore, dimensionless: $\beta / \alpha =  ( \xi -1)/(\pi \xi) $.
In \cite{aw} , the corresponding quantum BO equation has terms 
with different dimensions to (\ref{hameqm16}) and one coupling constant,
which is given by $(\xi -1)/(2\sqrt{\xi})$ and that should be put in 
correspondence with $( \xi -1)/(\pi \xi )$.
Note, however, that in \cite{aw} , the
Kac-Moody algebra has no $\xi$ factor as in (\ref{walg1}), so that one needs to introduce it for 
a correct comparison. The coupling $\alpha$ may be eliminated by a redefinition of the time coordinate.
We therefore find a complete agreement between the dynamics of the density field predicted by quantum
hydrodynamics and that of quantum field theory.
A related study on the classical hydrodynamics of the CS model and the (classical) Benjamin-One equation
\cite{boreview} is available in \cite{six} .

Finally, we discuss a physical interpretation of (\ref{wdecomp}): the Hilbert transform associates to 
a quantum field decomposed in Laurent modes
another one. This mapping is non-local in time, and involves 
both the distant past and future with respect to the field at present time. The field is a function 
of the coordinate $z$. In radial quantization, a growing $|z|$ amounts to time evolution, from the 
distant past ($|z| \to 0$) to the distant future ( $|z| \to \infty$ ). Laurent expansion can be viewed 
as a partial wave decomposition. In this setting, the Hilbert transform may
be viewed as transformation that takes an instantaneous function (potential) to a retarded or advanced one, as in the picture
of electromagnetic charge wave emission. This idea was developed by Wheeler and Feynman \cite{wheefeyn} and 
it means that the terms with negative sign in (\ref{wdecomp}) are interpreted as circular waves moving backwards in time 
from the distant future, 'the advanced waves' . Note that  this interpretation is also consistent with the classical
interpretation of the surface waves described by the BO equation, involving deep water and therefore long time
reflected waves in the bottom of the sea. 

We have explicitly shown how the Hilbert transform acting on the density 
field performs the separation of positive and negative Laurent modes, reversing the sign of half of them. 
Notice that the decomposition (\ref{wdecomp}) arises as an effect on the 
density field by the CS interaction term in the hamiltonian (\ref{hcsf2}).

\section{Conclusions} 

The EFT that describes the thermodynamic limit of the CS model may be cast as a 
$(c,{\overline c})=(1,1)$ CFTs with extended symmetry $\winf \times {\overline \winf}$ and chiral and
antichiral sectors that are isomorphic. The Hilbert space and partition
function for these theories are well known \cite{bpz,cdtz}. Nonetheless, what singles out this theory 
is the specific dynamics and time evolution induced by its effective hamiltonian. 
In particular, the $1/r^2$ interaction in the CS model implies the finiteness of the expansion
of the effective hamiltonian in terms of the $\winf$ generators (\ref{hcsf}) (see also \cite{lanlif}).

The Benjamin-Ono equation obeyed by the density field is not the only match 
between the descriptions based on quantum hydrodynamics and effective field
theories, since both 
the precise form of the hamiltonian (\ref{hcsf}) and the bosonic operator content (\ref{mod2})
agree as well. This equivalence may help to gain further insights into both  formulations. 
Alternatively, the knowledge of the time evolution of the density field rather than 
that of its vacuum expectation value clearly conveys further dynamical information. As a case in point, 
the time evolution of correlation functions involving the density may be greatly simplified 
by the use of its equation of motion. This constitutes a powerful tool to analyze the analytic
properties of the dynamical structure factor of the EFT of the CS model, which is a response function
to externally induced density fluctuations and is a relevant probe for analyzing the internal structure 
in condensed matter systems \cite{dsf}. On a more technical level, the role of projected density 
operators such as (\ref{halfw}) may be a useful resource elsewhere in CFTs studies.

Besides the lowest energy excitations, the algebraic structure of the $\winf$ algebra may also be used
to incorporate higher energy processes beyond its original domain of applicability. For instance, in the context of the 
quantum Hall effect, this idea
has proven to be successful in extending some of the edge states properties into the bulk
\cite{cama}. For the EFT of the CS model, this program should pursue a deeper understanding 
of the role of the Bogoliubov transformation and the high energy fermionic processes such as 
the backward scattering ones. In particular, the role of the chiral-antichiral interaction 
studied in \cite{boze3} and  its relation to the bosonic field duality 
of the compactification radius discussed after (\ref{rtil}).

We conclude by remarking that the consistent physical picture arising from
the continuous limit of the CS model is that of a 1D quantum compressible fluid, with non-linear waves of the
Benjamin-Ono type, that involves the two chiralities. This view stresses similarities and differences with the quantum
incompressible fluids that appear in the quantum Hall effect. We believe 
that this qualitative picture may be useful in future developments.

%
\def\NP{{\it Nucl. Phys.\ }}
\def\PRL{{\it Phys. Rev. Lett.\ }}
\def\PL{{\it Phys. Lett.\ }}
\def\PR{{\it Phys. Rev.\ }}
\def\IJMP{{\it Int. J. Mod. Phys.\ }}
\def\MPL{{\it Mod. Phys. Lett.\ }}

\end{document}